\documentclass[a4paper, 12pt]{article}
\usepackage{fullpage} 
\usepackage{bm} 
\usepackage{amsmath}
\usepackage{graphicx}
\usepackage{color}
\usepackage{url}
\usepackage[numbers]{natbib}

\begin{document}

\section*{Strong coupling and the C=O vibrational bond}

William L. Barnes\\

Department of Physics and Astronomy,

University of Exeter,

Exeter, EX4 4QL,

United Kingdom\\

email: w.l.barnes@exeter.ac.uk

\hfill \today

\section*{Abstract}

In this technical note we calculate the strength of the expected Rabi splitting for a molecular resonance. By way of an example we focus on the molecular resonance associated with the C=O bond, specifically the stretch resonance at $\sim$1730 cm$^{-1}$. This molecular resonance is common in a wide range of polymeric materials that are convenient for many experiments, because of the ease with which they may be spin cast to form optical micro-cavities, polymers include PVA~\cite{Shalabney_NatComm_2015_6_5981} and PMMA~\cite{Long_ACSPhot_2014_2_130,Menghrajani_ACSPhot_2019_6_2110}. Two different approaches to modelling the expected extent of the coupling are examined, and the results compared with data from experiments. The approach adopted here indicates how material parameters may be used to assess the potential of a material to exhibit strong coupling, and also enable other useful parameters to be derived, including the molecular dipole moment and the vacuum cavity field strength.\\

\section{Introduction}
\label{sec:intro}

The strength of the interaction of N molecular resonators and a cavity mode, $g_N$, is based on the electric dipole interaction. We consider molecular resonances that involve electric dipole transitions, at angular frequency $\omega_0$ and that have a transition dipole moment $\mu$. It is the interaction of this dipole moment with the cavity vacuum field, $E_{\textrm{vac}}$ that we are interested in, no external source of light is involved~\cite{Rider_CP_2021_62_217}. The interaction energy for our electric dipole is given by $\mu.E_{\textrm{vac}}$ (where we have assumed that the dipole moment and field are aligned). The (RMS) strength of the vacuum field is $\sqrt{\hbar\omega_0/2V\varepsilon_0\varepsilon_b}$, where $\varepsilon_b$ is the background permittivity of the molecular material, and $V$ is the volume of the cavity mode~\citep{Rider_CP_2021_62_217}. To make contact with the literature it is convenient to write the interaction energy for N molecules in a cavity as $\hbar \Omega_R$ and, since we often wish to know how far from the original molecular resonance energy $\omega_0$ the two polaritons $\omega_{\pm}$ are, we write $\Omega_R=\omega_+-\omega_-=2g_N$. We need one extra piece of information, the interaction energy scales as the square root of the number of dipoles (molecules) involved~\cite{Rider_CP_2021_62_217}, i.e. it scales with $\sqrt{N}$. We thus have,

\begin{align}
\Omega_R &= 2 g_N = \frac{2}{\hbar}\sqrt{N} \mu\,E_{\textrm{vac}} = \frac{2}{\hbar}\sqrt{N} \mu\,
\sqrt{\frac{\hbar \omega_0}{2V_m\,\varepsilon_0\,\varepsilon_b}}
\label{eq:Rabi-1}
\end{align}

\noindent so that the material parameters we need to calculate the Rabi splitting are the dipole moment and frequency associated with the molecular transition, $\mu,\,\omega_0$, the concentration of the molecular resonators, $\sqrt{N/V}$, and the background permittivity of the host in which the molecules are embedded, $\varepsilon_b$; note we have assumed that the molecules fill the volume of the cavity mode. In addition, if we will probably want to see whether the coupling strength dominates over the molecular and cavity (de-phasing) decay rates ($\gamma_M$ and $\gamma_C$ respectively), i.e. whether,

\begin{equation}
\label{eq:criterion-1}
g_N>\gamma_C,\gamma_M,
\end{equation}

\noindent so that we will need to determine $\gamma_M$ and $\gamma_C$. Out of curiosity, we may also want to evaluate the mode volume $V$, the quality factors of the molecular and optical resonances, $Q_M$ and $Q_C$ respectively, and the vacuum field strength $E_{vac}$. Looking at equation \eqref{eq:Rabi-1} we can see that to evaluate the interaction strength $g_N$ we will need to determine: the dipole moment $\mu$; the resonance frequency $\omega_0$; the concentration $N/V$ of molecular resonators, and and the refractive index of the host medium, $n_b$. Below we look at two ways of accomplishing this.\\

\section{The parameters we need}
\label{sec:params}

Let us look at the essentials first, i.e. the number density of C=O bonds, the C=O bond stretch resonance frequency, and the associated dipole moment. From easiest to hardest these are:

\subsection{Resonance frequency, $\omega_0$.} 
\label{sec:resonance frequency}

The resonance frequency can easily be determined from IR transmission measurements, and is found to be~\cite{Menghrajani_ACSPhot_2019_6_2110} $\sim$ 1734 cm$^{-1}$, which is equivalent to a wavelength of 5.8 $\mu$m, and an angular frequency of 3.26$\times$10$^{14}$ rad s$^{-1}$. We will look at the IR transmission spectrum in more detail below.

\subsection{Number density, $\sqrt{N/V}$.}
\label{sec:number density}

Shalabney et al. use the polymer PVA$_{\rm{C}}$, for which the density is~\cite{polymerdatabase} 1.19 g cm$^{-3}$. Thus, 1 cm$^3$ contains 1.19 g of the polymer. The molecular weight of PVA$_{\rm{C}}$ is 86~\cite{polymerdatabase}, so that 1 cm$^3$ contains $1.19/86$ moles. Since each repeat unit contains one C=O bond, the number of C=O bonds per cm$^3$ is thus $8.33\times10^{21}$, so that the density of bonds is $8.33\times10^{27}$ per m$^3$. We assume these numbers also apply to the spun films used in the strong coupling experiments.

\subsection{Dipole moment, $\mu$.}
\label{sec:dipole moment}

\subsubsection{dipole moment from Lorentz oscillator model}
\label{sec:dipole moment LO}

\begin{figure}[h!]
\centering
\includegraphics[width=0.90\linewidth]{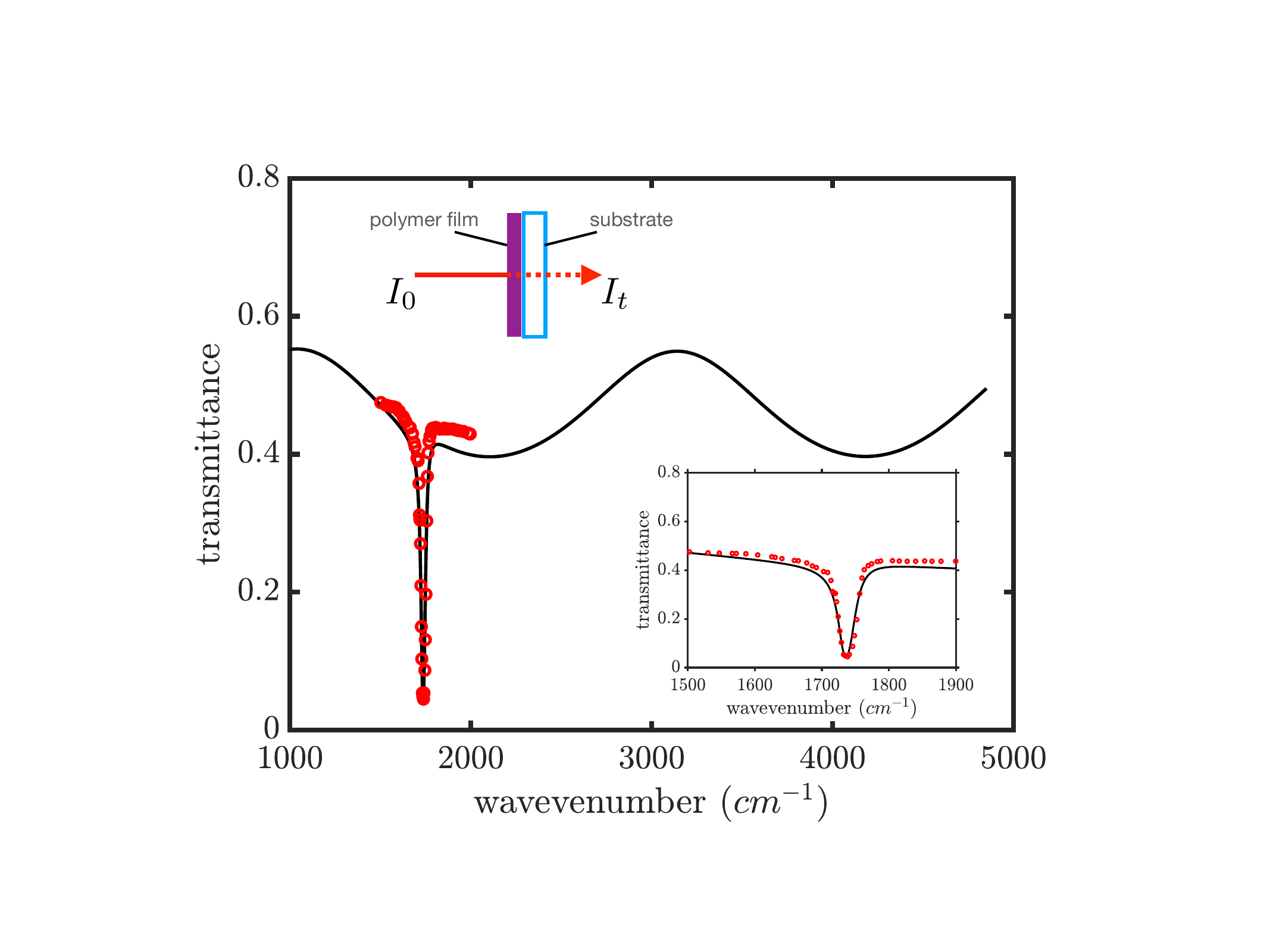}
\caption{\textbf{IR transmittance spectrum of PVA$_{\rm{C}}$}. Shown as red points are data transcribed from the report of Shalabney {\textit{et al.}}~\citep{Shalabney_NatComm_2015_6_5981}. Shown as a black line is the result of a Fresnel-based match to the experimental data. The inset lower right shows the spectral region around the C=O resonance in more detail. The inset upper left shows the configuration of the transmittance measurement, the PVA$_{\rm{C}}$ film is on a Ge substrate. See table [\ref{table:osc_strengths}]
 for details of the parameters used.}
\label{fig:IR_trans}
\end{figure}

The first approach to evaluate the dipole moment we use here is to fit a Lorentz oscillator (LO) model for the permittivity of the PVA$_{\rm{C}}$ to the measured transmission spectrum. The LO model is incorporated into a Fresnel-based calculation of the transmittance, the sample consisting of the spun polymer film on top of a substrate, see fig~\ref{fig:IR_trans}. A convenient `know nothing in advance' formulation of the Lorentz oscillator permittivity is,

\begin{equation}
\label{eq:LO_eps_1}
\varepsilon(\omega) = \varepsilon_{\rm{b}}+\frac{{f'}\omega_0^2}{\omega_0^2-\omega^2-i\gamma_{MD}\omega},
\end{equation}

\noindent where $\varepsilon(\omega)$ is the frequency dependent permittivity, $\varepsilon_{\rm{b}}$ is a constant across the frequency range of interest (it represents the contribution to the permittivity of resonances in other spectral regions). The molecular damping rate is $\gamma_{MD}$ ($\gamma_{MD}$ corresponds to the full width at half maximum of the resonance), note that the de-phasing rate and the damping rate are related through, $\gamma_M=\gamma_{MD}/2$. The reduced oscillator strength of the transition is $f'$.
For the Fresnel calculation of the transmittance we also need the thickness of the polymer layer, and the refractive index of the substrate. The dipole moment $\mu$ is related to the oscillator strength, $f$, by~\cite{NandH},

\begin{equation}
\label{eq:Kuhn_1}
f = \frac{2\, m_e\, \omega_0}{3\,\hbar\, e^2}|\mu|^2,
\end{equation}

\noindent so that if we can find $f$ then we can use \eqref{eq:Kuhn_1} to find the dipole moment. (The difference between $f'$ and $f$ is discussed below.)\\ 

The experimental information we have to work with is the IR transmission spectrum. Shalabney {\textit{et al.}} provide such a spectrum for a film of PVA$_{\rm{C}}$ on a Germanium substrate (for which we take the permittivity to be $\varepsilon_{\rm{Ge}}$=16.0, the superstrate is air). In assessing the transmittance data we need to take account of the reflection that occurs at the substrate/air interface, to simulate the experimental data we thus need to multiply our Fresnel-calculated data by 0.62 to take account of the transmittance of this interface \footnote{Note that to normalize their data Shalabney {\textit{et al.}} use a bare Ge substrate. This gives two air/Ge interfaces rather than one polymer/Ge interface and one Ge/air interface. There may also be some scattering, e.g. if the Ge was not polished on both sides. The result is that their data need to be scaled somewhat, a rare instance of adjusting the data rather than the model. The formula used for the Transmission of the Ge/air interface is $T=1-R=1-((n_{\rm{Ge}}-1)/(n_{\rm{Ge}}+1))^2=0.64$}. In the experiment the overall transmittance of the sample was measured. Figure \ref{fig:IR_trans} shows the transcribed experimental data from Shalabney {\textit{et al.}} (red data points) together with the result of a Fresnel-based calculation for the transmittance, where the parameters of the Lorentz oscillator model for the C=O bond have been varied to provide a reasonable `by eye' fit to the experimental data. The inset shows the spectra in the neighbourhood of the C=O stretch, the main part of the figure covers a wider energy range. In the main figure there is a gentle periodic modulation of transmittance, arising from interference due to the two surfaces of the polymer film, which are separated by $d=1.70\,\mu m$, see fig 1a in~\cite{Shalabney_NatComm_2015_6_5981}. The parameters used in the model are given in table table \ref{table:osc_strengths} below.\\

Now, although equation \eqref{eq:LO_eps_1} is convenient, it is not appropriate to use in finding the dipole moment because $f'$ does not match the physical origin of the Lorentz oscillator model correctly. Instead we should use,

\begin{equation}
\label{eq:LO_eps_2}
\varepsilon(\omega) = \varepsilon_{\rm{b}}+\frac{{f}\omega_{\rm{p}}^2}{\omega_0^2-\omega^2-i\gamma_{MD}\omega},
\end{equation}

\noindent where,

\begin{equation}
\label{eq:plasma_freq}
\omega_{\rm{p}} = \sqrt{\frac{N\,e^2}{V\varepsilon_0\,m_{\rm{e}}}},
\end{equation}

\noindent with $e$ the electronic charge and $m_{\rm{e}}$ the electron mass. Here $N/V$ is the number density of C=O bonds. Note that $\omega_{\rm{p}}$ is \textbf{\textit{not}} a plasma frequency, there is no plasma in the PVA$_{\rm{C}}$, rather we should think of it as short hand for the \textit{rhs} of equation \eqref{eq:plasma_freq}. Shalabney \textit{et al.} use a third version,

\begin{equation}
\label{eq:LO_eps_3}
\varepsilon(\omega) = \varepsilon_{\rm{b}}+\frac{f_k}{k_0^2-k^2-i\gamma_{MDk}\, k},
\end{equation}

\noindent where now the parameters are given in terms of wavenumber, ($\rm{cm}^{-1}$).\\

We will now find $f$ and $\gamma$ and will also find $f_k$ and $\gamma_{MDk}$, so that we may make a comparison between our results and those of Shalabney \textit{et al.}.  Comparing equations \eqref{eq:LO_eps_1} and \eqref{eq:LO_eps_2} we have,

\begin{equation}
\label{eq:LO_freq_comp}
f'\omega_0^2 = f \omega_{\rm{p}}^2.
\end{equation}

\noindent From the `fit' shown in figure \ref{fig:IR_trans} we note that we used $\omega_0=3.26\times 10^{14}$ rad s$^{-1}$ and $f'=0.018$, and from this we can calculate $\omega_{\rm{p}}$ via \eqref{eq:plasma_freq} using the bond density discussed earlier, i.e., $N/V=8.0 \times 10^{27}$ m$^{-3}$. Doing so we find $\omega_{\rm{p}}=5.1\times 10^{15}$, so that $f=0.7 \times 10^{-4}$. The background permittivity was taken to be $\varepsilon_b=1.99$~\cite{Shalabney_NatComm_2015_6_5981}. Next, noting that to convert rad s$^{-1}$ to cm$^{-1}$ we divide by $1.88 \times 10^{11}$, we find, $f_k = 54 \times 10^3$, $k_o=1734$ $cm^{-1}$, and $\gamma_{MDk}=16$ $cm^{-1}$. In table \ref{table:osc_strengths} we bring all of these data together, including the data from Shalabney \textit{et al.}, i.e. their results from fitting their own data. To a reasonable approximation the results from the modelling described above agree with those of Shalabney \textit{et al.}\\

\begin{table}[ht!]
\centering

\begin{tabular}{ |p{3.5cm}|p{3.5cm}|p{3.5cm}|p{3.5cm}|}
 \hline
 \hline
  equation~\eqref{eq:LO_eps_1}\newline (rad s$^{-1}$)& equation~\eqref{eq:LO_eps_2}\newline (rad s$^{-1}$)& equation~\eqref{eq:LO_eps_3}\newline (cm$^{-1}$)& Shalabney \textit{et al.}\newline (cm$^{-1}$) \\

\hline
 &  & &\\
$\omega_0=3.26\times 10^{14}$ & $\omega_0=3.26\times 10^{14}$ & $k_0=1734$ & $k_0=1739$\\
 &  & &\\
 & $\omega_{\rm{p}}=5.1\times 10^{15}$ & &\\
 &  & &\\
\hline
 &  & &\\
$\gamma_{MD}=3.0\times 10^{12}$ & $\gamma_{MD}=3.0\times 10^{12}$ & $\gamma_{MDk}=16$ & $\gamma_{MDk}=13$\\
 &  & &\\
\hline
 &  & &\\
$f'=0.018$ & $f=7.0\times 10^{-5}$ & $f_k=54\times10^3$ & $f_k=50\times10^3$\\
 &  & &\\
\hline
\hline
\end{tabular}
\caption{\textbf{Table of Lorentz oscillator parameters for the C=O bond} for the different models considered. The oscillator strengths are dimensionless, the units for the other parameters are as given at the head of each column.}
\label{table:osc_strengths}
\end{table}

Now let us use equation \eqref{eq:LO_eps_1} to find the dipole moment, we have from equation \eqref{eq:Kuhn_1},

\begin{equation}
\label{eq:dip_mom_1}
|\mu|^2 = \frac{3\,\hbar\, e^2}{2\, m_e\, \omega_0}f.
\end{equation}

\noindent Introducing the numerical values we find $|\mu|=0.97\times 10^{-30}$ Cm = 0.29 D [where 1 Debeye (D) = $3.3 \times 10^{-30}$ Cm]. This is in the range of vaules given by Grechko and Zanni for various materials~\cite{Grechko_JCP_2012_137_184202}, note that Shalabney \textit{et al.} state a value for the dipole moment of 1D.


\subsubsection{Dipole moment from molar absorption model}
\label{sec:dipole moment EX}

Chemists frequently make use of extinction measurements, just as we have done above. However, rather than trying to model the extinction (or equivalently the transmittance),  chemists usually adopt a different approach based on the molar absorption coefficient, $\epsilon$. The optical measurement is usually carried out on a solution of the molecules of interest, of concentration $C_m$, held in a sample chamber (cuvette) that provides a path length $l$. The \textbf{molar absorption coefficient}\footnote{Note that this is often called the molar extinction coefficient, but molar absorption coefficient is the correct term, see~\cite{Braslavsky_PAC_2007_79_293}.} is given by~\cite{Turro,Atkins_and_Freidman_11},

\begin{equation}
\label{eq:mol_abs_coeff}
\epsilon = \rm{log}_{10}\left(\frac{I_0}{I_{\rm{t}}}\right)\frac{1}{\textit{l}\,C_m}.
\end{equation}

\noindent $I_0$ and $I_{\rm{t}}$ are the incident and transmitted intensities, and we can write for the transmittance $T$, $I_0/I_{\rm{t}}=1/T$. As ever in calculating numerical quantities, using the correct units is vital. For a physicist SI units seem obvious, but they are not the ones that chemists use here. Instead, the units for the path length, $l$, are cm, whilst the units for the molecular concentration, $C_m$, are moles per dcm$^{-3}$, i.e. moles per litre. Turro(see~\cite{Turro} equation 5.10) and Valeur and Berberan-Santos (see~\cite{MolFluo} equation 2) give the oscillator strength as\footnote{Note that~\cite{Turro} uses wavenumber (cm$^{-1}$), whilst~\cite{MolFluo} uses Hz, we use the wavenumber (cm$^{-1}$) version here since this is convenient in the infrared},

\begin{equation}
\label{eq:mol_abs_coeff_1}
f = 4.32 \times 10^{-9}\int\epsilon(\bar{\nu})\, d\bar{\nu},
\end{equation}

\noindent where $\bar{\nu}$ is wavenumber in cm$^{-1}$ and, using~\cite{Turro} equation 5.40, this can often be approximated as,

\begin{equation}
\label{eq:mol_abs_coeff_2}
f \approx 4.32 \times 10^{-9}\epsilon_{\rm{max}}\, \delta\bar{\nu},
\end{equation}

\noindent where $\delta\bar{\nu}$ is the width (FWHM) of the extinction feature in wavenumbers (cm$^{-1}$) \footnote{For an alternative derivation based on the Einstein coefficients, see~\cite{Methods_3}.}. We note that the oscillator strength found using equations~\eqref{eq:mol_abs_coeff_1} and~\eqref{eq:mol_abs_coeff_1} is less than the oscillator strength employed in the Lorentz oscillator model, by a factor of $n_b$. This is associated with the change in energy density of the light in the molecular material (\textit{c.f.} air), see equation 9.29 in~\citep{Kuhn_Forsterling_Waldeck}. As a result the oscillator strength derived from extinction-type measurements should instead be written as,

\begin{equation}
\label{eq:mol_abs_coeff_3}
\frac{f}{n_b} = 4.32 \times 10^{-9}\int\epsilon(\bar{\nu})\, d\bar{\nu} \approx 4.32 \times 10^{-9}\epsilon_{\rm{max}}\, \delta\bar{\nu}.
\end{equation}

For the C=O data of Shalabney \textit{et al.} (see figure \ref{fig:IR_trans}), $\epsilon_{\rm{max}}$ corresponds to $T=0.045/0.45=1/10$, and the path length is $l=1.7 \times 10^{-4}$ cm. For the molecular concentration, $C_m$, 1 cm$^3$ contains 1.19/86 moles, so that there are 13.5 mol dm$^{-3}$, \textit{i.e.}  $C_m=$13.5 mol dm$^{-3}$. Bringing these factors together we can use \eqref{eq:mol_abs_coeff} to evaluate $\epsilon_{\rm{max}}$ and find $\epsilon_{\rm{max}}=436$. Finally we can use  $\delta\bar{\nu}\,\approx$ 30 cm$^{-1}$ (estimated from figure~\ref{fig:IR_trans_2}) to calculate $f$ via equation~\eqref{eq:mol_abs_coeff_3}, we find $f=7.3\times10^{-5}$. This compares with the $7.0\times10^{-5}$ we obtained earlier, see table \ref{table:osc_strengths}, surprisingly similar given the rather crude approximations we have made. The associated dipole moment is $\mu$ = 0.30 D.\\

\section{The `derived' parameters}
\label{sec:derived params}

\subsection{the coupling strength}
\label{sec:coupling strength}

We can now look at the coupling strength and compare it to the decay rates to see if the strong coupling regime applies. From equation \eqref{eq:Rabi-1} we can calculate the value of $g_N$ for the C=O bond in PVA$_{\rm{C}}$, doing so we find $g_N=$81 cm$^{-1}$ using parameters from the LO model (section~\ref{sec:dipole moment LO}), and $g_N=$82 cm$^{-1}$ using parameters directly from the extinction data (section~\ref{sec:dipole moment EX}). Note that in using equation \eqref{eq:Rabi-1} we need to divide the right-hand side by $\sqrt{3}$, see footnote \footnote{We divide the right-hand side by $\sqrt{3}$ since here we are assuming that the dipole moments associated with all of the C=O bonds are randomly alighted with respect to the cavity field, i.e. that the C=O bonds are randomly oriented in the spun polymer film. This difference between random and aligned is the origin of the factor of 3 in the denominator of \eqref{eq:Kuhn_1}.}.

\subsection{the strong coupling condition}
\label{sec:strong coupling condition}

Recall that for strong coupling we require the ensemble coupling strength, $g_N$, to be greater than the molecular and cavity de-phasing rates, \textit{i.e.} we require $g_N>\gamma_M,\gamma_C$. From table \ref{table:osc_strengths} we see that, in cm$^{-1}$, the molecular de-phasing rate is $\gamma_{Mk}=\gamma_{MDk}/2$ = 8 cm$^{-1}$. For the cavity resonance Shalabney \textit{et al.} give the damping rate via the FWHM of their cavity resonance, as measured in transmission, to be 140 cm$^{-1}$ (17 meV), so that $\gamma_{Ck}=\gamma_{CDk}/2$ = 70 cm$^{-1}$. Thus the coupling strength, $g_N$ $\sim$80 cm$^{-1}$, exceeds the damping rates, confirming that the strong coupling regime has been reached. It is interesting to note that the measured Rabi splitting in the experiment of Shalbney \textit{et al.} is $\sim$170 cm$^{-1}$, implying a value for $g_N$ of $\sim$85 cm$^{-1}$.

\section{Mode widths}
\label{sec:mode widths}

\subsection{A cautionary tale}
\label{sec:a cautionary tale}

\begin{figure}[h!]
\centering
\includegraphics[width=0.90\linewidth]{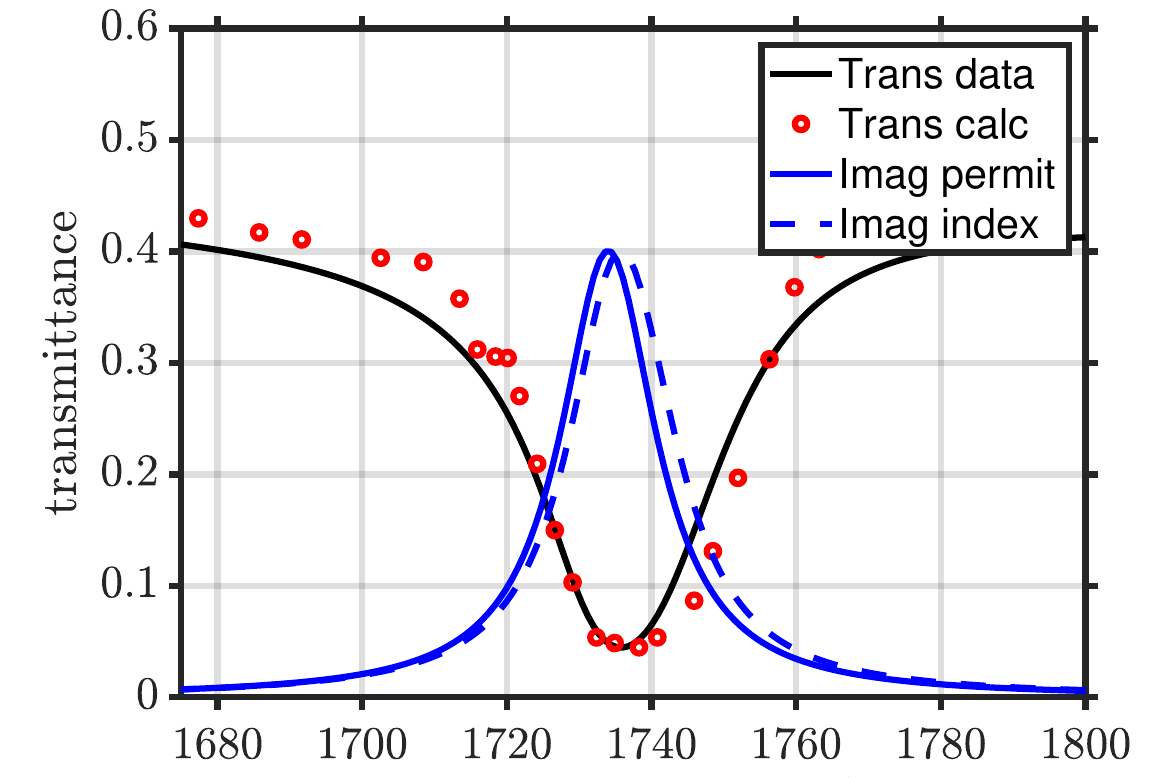}
\caption{\textbf{IR transmittance spectrum of PVA and material (PVA) response$_{\rm{C}}$}. Shown as red points are data transcribed from the report of Shalabney {\textit{et al.}}~\citep{Shalabney_NatComm_2015_6_5981}. Shown as a black line is the result of a Fresnel-based match to the experimental data. Also shown (scaled) are the imaginary parts of the PVA permittivity and the PVA refractive index.}
\label{fig:IR_trans_2}
\end{figure}

An element of confusion was encountered in carrying out some of the analysis described here that may have  implications for looking at a range of published strong coupling data. The problem has to do with the width of the transmittance minimum, it is not the same as the width that goes into the Lorentz oscillator model. This is illustrated in Figure~\ref{fig:IR_trans_2} where a closer zoom-in of the transmittance data shown in figure~\ref{fig:IR_trans} is given. In addition, the imaginary part of the PVA permittivity, and the imaginary part of the PVA index have been added. The imaginary part of the PVA permittivity has a FWHM of 16 cm$^{-1}$, whilst the transmittance dip has a FWHM, estimated from figure~\ref{fig:IR_trans_2}, of nearly 30 cm$^{-1}$. Taking the measured transmittance width as the FWHM to be used in the Lorentz oscillator model is a mistake, the measured transmittance width is instead the width that is needed when making use of the extinction data to evaluate the oscillator strength. Reader beware!\\

\subsection{Mode widths and Q factors}
\label{sec:q factors}

For completeness, let us calculate the relevant Q factors. For the molecular resonance this is $Q_M=\omega_0/\gamma_{MD}$ = 1734/16 cm$^{-1}$ = 108. For the cavity resonance it is $Q_M=\omega_0/\gamma_{CD}$ = 1734/140 cm$^{-1}$ = 12.

\subsection{Cavity volume, vacuum field strength, and number of molecules coupled}
\label{sec:vac field}

It may also be useful to look at the vacuum field strength. A reasonable approximation can be found by estimating the mode volume $V$, and using the well-known experssion for the vacuum field strength~\cite{Rider_CP_2021_62_217},

\begin{equation}
\label{eq:vac_field}
E_{\rm{RMS}}=\sqrt{\frac{\hbar\omega_0}{2\,\varepsilon_0\,\varepsilon_b V}}.
\end{equation}

For the mode volume, to a reasonable approximation this is $d\pi r_{\rm{eff}}^2$ where $d$ is the cavity thickness, and $r_{\rm{eff}}$ is the radius of the mode, this in turn is determined by the width of the mode in terms of in-plane wavevector~
\cite{Bjork_PRA_1993_47_4451,Kai_JJAP_2002_41_7398} as,

\begin{equation}
\label{eq:cavity radius}
r_{\rm{eff}}=\frac{\lambda}{4\,n_b}\frac{\pi \sqrt{R}}{1-R},
\end{equation}

\noindent where $R$ is the reflectance (intensity reflection coefficient) of the cavity mirrors. A Fresnel multilayer calculation can be used to work out $R$ for the upper and lower mirrors. The mirrors are 10 nm thick gold (The permittivity of gold at this wavelength was estimated to be $\varepsilon_{\rm{Au}}$ = -1000 + 200i, based on data in ~\cite{LandH_Palik}). For the upper mirror the three media are PVA/Au/air, for the lower mirror they are PVA/Au/Ge. Values of $R_{\rm{upper}}$ = 0.85 and $R_{\rm{lower}}$ = 0.8 were found. Taking an average value of $\sim$ 0.8 we find $r_{\rm{eff}}$ = 7.3 $\mu$m so that the mode volume is $V=d\pi r_{\rm{eff}}^2$ = 300 $\mu$m$^3$. We can now calculate the vacuum field strength with the help of equation \eqref{eq:vac_field}, we find the field strength to be $E_{\rm{RMS}}$ = 1.7 $\times 10^3$ Vm$^{-1}$. (Note that Shalabney \textit{et al.} took their mode volume to be $V=(\lambda/n_b)^3$ = 70 $\mu$m$^3$).\\

Finally we can estimate the number of C=O bonds involved from the density (8.3 $\times$ 10$^{27}$ m$^{-3}$) and the mode volume (300 $\times 10^{-18}$ m$^3$) as N = 2 $\times$ 10$^{12}$. It is now clear that the single molecule coupling strength, $g_N/\sqrt{N}$ = 1 $\times 10^{-4}$ cm$^{-1}$ is much much smaller than the de-phasing rates, strong coupling here really is a collective effect.

\section*{Conclusion}
We have shown how two relatively simple models can be used to understand the extent of the Rabi-splitting observed in vibrational strong coupling experiments. The analysis has been presented in a somewhat tutorial style with the aim of giving those entering the field an easy entry point in trying to link the extent of the Rabi-splitting with bulk material parameters. In addition, the analysis presented here allows a number of related parameters to be evaluated, notably the dipole moment, the oscillator strength, the vacuum field strength and the cavity mode volume of planar Fabry-Perot-type cavities. 

\section*{Acknowledgements}
The author is grateful to Marie Rider and Kishan Menghrajani for valuable discussions and to the European Research Council for funding Project Photmat (ERC-2016-AdG- 742222: www.photmat.eu). Note that this study did not generate any new data.

\bibliographystyle{iopart-num}

\bibliography{vib}

\providecommand{\newblock}{}
\begin{thebibliography}{10}
\expandafter\ifx\csname url\endcsname\relax
  \def\url#1{{\tt #1}}\fi
\expandafter\ifx\csname urlprefix\endcsname\relax\def\urlprefix{URL }\fi
\providecommand{\eprint}[2][]{\url{#2}}

\bibitem{Shalabney_NatComm_2015_6_5981}
Shalabney A, George J, Hutchison J, Pupillo G, Genet C and Ebbesen T~W 2015
  {\em Nature Communications\/} {\bf 6} 1--6

\bibitem{Long_ACSPhot_2014_2_130}
Long J~P and Simpkins B~S 2015 {\em ACS Photonics\/} {\bf 2} 130--136

\bibitem{Menghrajani_ACSPhot_2019_6_2110}
Menghrajani K~S, Nash G~R and Barnes W~L 2019 {\em ACS Photonics\/} {\bf 6}
  2110--2116

\bibitem{Rider_CP_2021_62_217}
Rider M~S and Barnes W~L 2021 {\em Contemporary Physics\/} {\bf 62} 217--232

\bibitem{polymerdatabase}
 2018 {Poly(Vinyl Acetate)}
  \url{https://polymerdatabase.com/polymers/polyvinylacetate.html} [accessed
  02-December-2022]

\bibitem{NandH}
Novotny L and Hecht B 2006 {\em Principles of nano-optics\/} 1st ed (Cambridge,
  UK: Cambridge University Press)

\bibitem{Grechko_JCP_2012_137_184202}
Grechko M and Zanni M~T 2012 {\em The Journal of Chemical Physics\/} {\bf 137}
  184202

\bibitem{Braslavsky_PAC_2007_79_293}
Braslavsky S~E 2007 {\em Pure and Applied Chemistry\/} {\bf 79} 293--465

\bibitem{Turro}
Turro N~J 1991 {\em Modern Molecular Photochemsitry\/} (Sausulito, California:
  University Science Books)

\bibitem{Atkins_and_Freidman_11}
Atkins P and Frediman R 2005 {\em Molecular Quantum Mechanics\/} 4th ed (Oxford
  University Press) ISBN 0-19--927498--3

\bibitem{MolFluo}
Valeur B and Berberan-Santos M~N 2012 {\em Molecular Fluorescence: Principles
  and Applications\/} 2nd ed (Weinheim: Wiley VCH)

\bibitem{Methods_3}
Weatherly T~L and Williams Q 1962 {\em Methods of Experimental Physics -
  Molecular Spectroscopy\/} vol~3 (New York: Academic Press)

\bibitem{Kuhn_Forsterling_Waldeck}
Kuhn H, F{\"o}rsterling H~D and Waldeck D~H 2009 {\em Principles of Physical
  Chemistry\/} 2nd ed (New Jersey: Wiley) ISBN 978-0-470-08964-4

\bibitem{Bjork_PRA_1993_47_4451}
Bj\"ork G, Heitmann H and Yamamoto Y 1993 {\em Phys. Rev. A\/} {\bf 47}
  4451--4463

\bibitem{Kai_JJAP_2002_41_7398}
Kai K, Oh-Hara C, Inoue H, Yamanaka T and Ujihara K 2002 {\em Japanese Journal
  of Applied Physics\/} {\bf 41} 7398

\bibitem{LandH_Palik}
Lynch D~W and Hunter W~R 1985 {\em Handbook of Optical Constants of Solids\/}
  (Academic Press Inc. (Orlando)) chap Optical constants of metals, pp 275--406

\end{thebibliography}

\end{document}